\documentclass[letterpaper]{article}
\pdfoutput=1
\usepackage{aaai}
\usepackage{times}
\usepackage{helvet}
\usepackage{courier}
\usepackage{graphicx}

\begin{document}
\title{Mutual Emotion-Cognition Dynamics}
\author{Mikhail I. Rabinovich \and Mehmet K. Muezzinoglu\\
BioCircuits Institute\\
University of California, San Diego\\
9500 Gilman Dr., La Jolla, CA, 92093-0402\\
}

\maketitle

\begin{abstract}
\begin{quote}
We present a new paradigm in the study of brain mental dynamics on the basis of the stable transient activity neural networks observed in experiments. This new approach is in contrast to traditional system analysis usually adopted in cognitive modeling. Transient dynamics offers a sound formalism of the observed qualities of brain activity, while providing a rigorous set of analysis tools. Transients have two main features: First, they are resistant to noise, and reliable even in the face of small variations in initial conditions; the sequence of states visited by the system (its trajectory), is thus structurally stable. Second, the transients are input-specific, and thus convey information about what caused them in the first place. This new dynamical view manifests a rigorous explanation of how perception, cognition, emotion, and other mental processes evolve as a sequence of activity patterns in the brain, and, most importantly, how they interfere with each other. The ideas discussed and demonstrated here lead to the creation of a quantitative theory of the human mental activity and can be deployed on artificial agents.
\end{quote}
\end{abstract}

\section{Introduction}
It looks like the popular opinion that ``it is much easier to invent an artificial brain than to discover the working principles of the human brain'' is true. This is because an artificial intellect only requires the execution of mental functions, and not the underlying brain mechanisms. This kind of modeling does not require the discovery of one of nature's greatest secrets, namely, {\it the origin of thoughts}.

A specific intellectual function cen be modeled in many different ways, and none of these ways need to be bio-mimetic. The success of artificial intelligence in discovering these ways in its own path is important for understanding the principles of the brain, because it provides us with a vision of how a real-life problem can be solved in general, but not necessarily of whether nature uses these principles. 

It is crucial to realize that the transfer of ``artificial ideas'' to the human brain may lead us in the wrong direction. Here are two examples: The first one is a traditional paradigm suggesting that the human brain is similar to a computer with a huge number of computing units. In contrast with the computer, however, which reorganizes incoming information, {\it the brain is an active system that is also able to generate new information by itself.} The brain is even able to generate its own emotions. 

The second example is related to emotions: Due to the profound interest in human-robot interactions, a new field has emerged that involves the creation of emotional robots \cite{FeA:05}. Modeling the emotion-cognition tandem of human mental life in robotic applications is a very promising direction that could qualitatively leverage the artificial intelligence. The state-of-the-art in emotional robot design is faithful to the algorithmic perspective, prescribes the solution in a sequential order. As a result the artificial brain solves cognitive and emotional problems step by step. However, both the cognition and the emotion are complex dynamical processes that involve the representation and the assessment of stimuli, stressors, and a prediction of possible coping behaviors. To carry out these, the human brain follows a {\it parallel}, {\it competitive}, and {\it dynamical} procedure in {\it continuous time}. We claim that integration of these priciples in the design is vital for the long-term success of an emotionally- and cognitively-capable artificial agent, and even for modeling disorders.

This paper accounts for these principles in an ecological framework and formulates a joint dynamical model for cognition and emotion. The model can be directly realized using basic numerical methods for real-time applications. In the following section, we revisit the short history of the dynamical approaches to the modeling of the brain's mental functions of emotion, cognition, perception, and consciousness. In Section III, we describe the model and a transient dynamical regime within its repertoire that is critical in our development. We provide some numerical examples in Section IV, followed by a brief discussion and concluding remarks in Section V.

\section{Dynamics of Cognition and Emotion}
\subsection{Background and Motivation}
As the two main players of human mental life, emotion and cognition have been under the spotlight for researchers for a long time both individually and jointly. The tradition for understanding thought based on dynamical systems theory has its roots in the cybernetics era of the 1940s.  It was a time when information theory, dynamics, and computation were brought together for studying the brain \cite{Ash:54}. However, with the dominance of symbolic artificial intellect, ``information-processing psychology,'' and the absence of a good experimental technology in the 1960s and 1970s, dynamical-systems-based approaches were not extensively pursued. More recently, the idea that dynamics is a relevant framework for understanding cognition has become popular again. For example, \cite{ThS:94} describe the development of kicking and reaching in infants in terms of dynamical notions such as the stability of attractors in a phase space that is defined by the body and environmental parameters. Movements to new stages in development are explained in terms of bifurcations to new attractors as a result of a change in the order parameters, e.g., infant weight, body length, etc., as the infant grows. Thelen and Smith believe that ``higher cognition'' is ultimately rooted in these types of spatial skills learned in infancy, and, thus, that higher cognition will itself be best understood dynamically. They contrast their account with traditional ``information processing'' theories of development, in which new developmental stages are caused by brain maturation and the increasing ability of maturing infants to reason logically.

\cite{PoG:95} have formulated the general idea that cognition should be characterized as a continual coupling among brain, body, and environment that unfolds in real time, as opposed to the discrete time steps of the artificial intellect. This is said to contrast with the computation's focus on ``internal structure,'' i.e., its concern with the static organization of information processing and representational structure in a cognitive system. A dynamical approach means that the organization of brain structures is also dynamical and functional, i.e., not only anatomical. Thus the dynamical approach to cognition is a confederation of research efforts bound together by the idea that natural cognition is a dynamical phenomenon and best understood in dynamical terms. This contrasts with the ``law of qualitative structure'' \cite{NeS:76} governing orthodox or ``classical'' cognitive science, which holds that cognition is a form of digital computation. 

The quest for the dynamical origin of emotions goes back many decades. \cite{Fra:35} devoted attention to the dynamics of emotions by describing emotional sequences together with their content. He emphasized that only the balance between incorporation, elimination and retention represents the fundamental dynamics of the biological process called life. Recently, \cite{Zau:03} has developed a two-dimensional approach in which both positive and negative emotions are conceptualized and measured as co-occurring simultaneous dynamical processes.


\subsection{The Natural Course of Mental Activity}
Although classical cognitive science has interpreted cognition as something {\it happening} over time, the dynamical approach sees cognition as {\it being} in time, i.e., as an inherently temporal phenomenon. For example, when a dynamical model of the information (sensory) coding is created, time is included in the coding space \cite{RVSA:06}. Details of timing (durations, rates, synchronizes, etc.) matter \cite{BuD:04}.

\begin{figure}[t!]
\centerline{\includegraphics[height=1.2in]{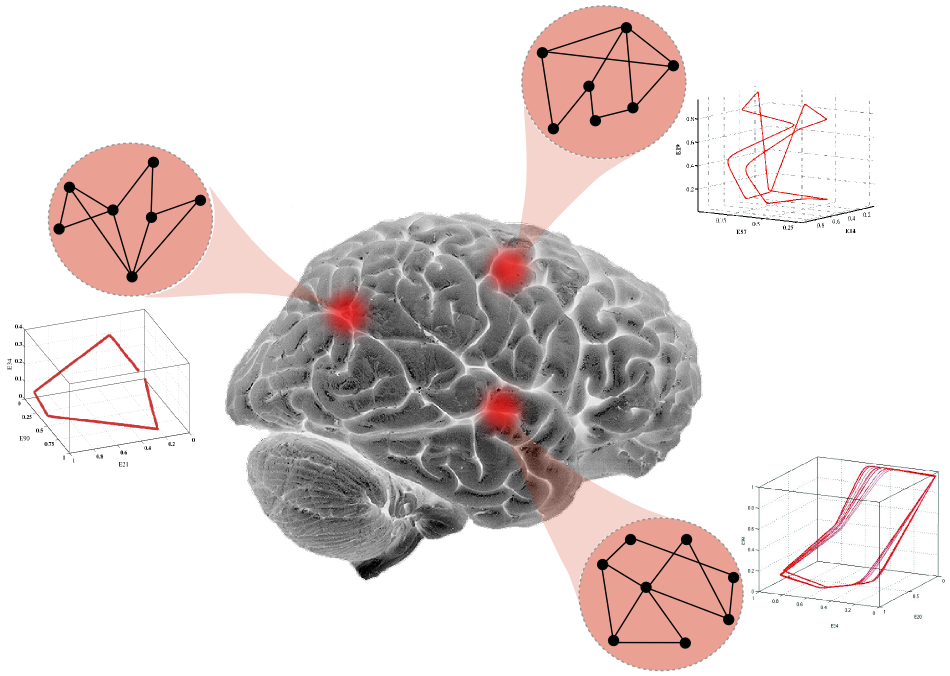}\hspace*{0.2cm}
\includegraphics[height=1.1in]{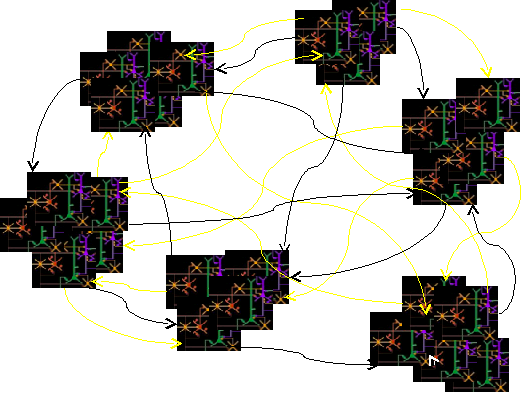}}
\caption{Brain centers are densely connected networks that can demonstrate a broad range of dynamical behaviors within their individual phase spaces, as illustrated on the left panel. Long-range (excitatory or inhibitory) connections couple these dynamics (as shown on right panel) that results in dynamical patterns observable on the joint phase space. }
\label{fig:hierarchy}
\end{figure}

New results in the brain imaging, particularly functional magnetic resonance imaging (fMRI) data, have revealed some fundamental properties and functional organization of the brain systems that correlate with emotion and cognitive functions \cite{PWT:02,DoM:06}. 

Each of the brain centers that form the functional emotional subcircuit or mode is itself a very complex dynamical system with several characteristic time scales (see Fig.~\ref{fig:hierarchy}). These systems are open to an enormous range of neural stimulations from a wide range of brain areas. The spatiotemporal pattern of brain activity underlying an emotion is typically very sensitive to the external or internal stimuli. Amygdala, for example, receives information from both cortical and subcortical structures. These include highly processed information from the visual system, the auditory cortex, the olfactory and gustatory neocortex and the somatosensory cortex. In short, it is directly informed about each of the five senses. The amygdala also receives projections from association cortex, from the thalamus (relaying basic, unprocessed sensory signals), the hippocampus (high level information about the relationship between objects and events in the external world), and from a range of structures that represent internal bodily states, such as hunger and thirst.

It is important to take into account that {\it emotions and cognition are active processes} that result in specific changing of the brain organization in time and dynamical brain's response to environmental information and representation of the self. These processes are determined by the functional (not necessary anatomical) connections between brain areas or neural circuits that participate in the execution of cognitive functions and generation of emotions. At different segments or steps of temporal emotional or cognitive process, these networks form different temporal sequences that execute and represent different emotions and cognitive functions in the brain.

Given the necessity that any useful analysis must be based on an accurate quantification of the investigated phenomenon, numerous attempts have been made to assess cognition and emotion. Being directly related to the processing of auxiliary information, cognition has attracted relatively more attention in these efforts, particularly in the form of task development assessing decision making tasks\cite{ScC:00,Hol:03}. Although, tests aiming solely at emotional quantities also exists (see, for example, \cite{PBB:85}), the assessment of emotions or their effects have often been attempted in a cognitive theme, involving, for example, appraisal \cite{Sch:93b,ThA:08}, decision making \cite{PeP:05}, or memory \cite{Lee:99,FBB:08}.

A dynamical system for the modeling of emotion, cognition and their interactions is a set of quantitative variables changing continually, concurrently, and inter-dependently over quantitative time in accordance with dynamical principles, which are embodied in a set of differential equations.

Based on an accurate quantification method, dynamics offers a very successful experience of dynamical modeling that scientists use to understand natural phenomena via nonlinear differential equations. This experience includes a set of concepts, proofs, and tools for understanding the behavior of systems in general. An important insight of dynamical systems theory is that behavior can be understood geometrically in some projection of the state (phase) space. The behavior can then be described in terms of attractors, transients, stability, coupling, bifurcations, chaos, etc.

\subsection{Metastability and Dynamical Principles}
We base our modeling effort on the following principles:
\begin{itemize}
\item{existence of metastable states representing the modes in the unified emotion-cognition working space,}
\item{structural stability of the transients that are formed by the switching of the system (brain) among these states,}
\item{ecological, i.e., competition principles governing these switchings.}
\end{itemize}

The first item, i.e., metastability, is a general nonlinear dynamics concept, which describes states of delicate equilibrium. Metastability in brain is a phenomenon, which is being studied in neuroscience to elucidate how the human mind process information and recognizes patterns. There are semi-transient signals in the brain, which persist for a while and are different than the usual equilibrium state \cite{ABG:95}. Metastability is a principle that describes the brain's ability to make sense out of seemingly random environmental cues \cite{OuK:06,Wer:07}. The metastable activity of the cortex can also be inferred from the behavior \cite{BrK:01}.
The metastability is supported by the flexibility of coupling among diverse brain centers or neuron groups \cite{Fri:97,Fri:00,INV:07,SMI:07,FiF:06}. The temporal order of the metastable states is determined by the functional connectivity of the underlying networks and their causality structure \cite{CBD:09}. 
The mathematical image of a metastable state is a saddle in the state space of the system. The image of the transition between these saddles is the unstable separatrix connecting them (see Fig.~\ref{fig:shc}). Such construction is named a heteroclinic chain. 



\subsection{Winnerless Competition as a Dynamical Mechanism of Transients Stability}
Competition without a winner (or, continuously changing winners), is a widely-known phenomenon in systems involving more than two interacting agents that satisfy a relationship similar to the voting paradox \cite{Bor:old,Saa:95} or the popular game rock-paper-scissors. 
The mathematical image of it is a heteroclinic  cycle, which was first formulated in \cite{BuH:80} in modeling convection in a rotating layer. 
As a generic dynamical phenomenon, which is rare in simple systems yet common in complex ones, the sequential switching among saddles can provide concise and constructive formulations in a variety of real-world problems \cite{ATHR:08}. Prototype dynamical models that are widely accepted in computational neuroscience \cite{WiC:73}, and ecology \cite{Lot:25} have been shown to exhibit a transitive winnerless competition for a fairly broad range of parameters \cite{RHA:06,HuR:04,ATHR:08}. 

\begin{figure}
\centerline{\includegraphics[height=1.0in]{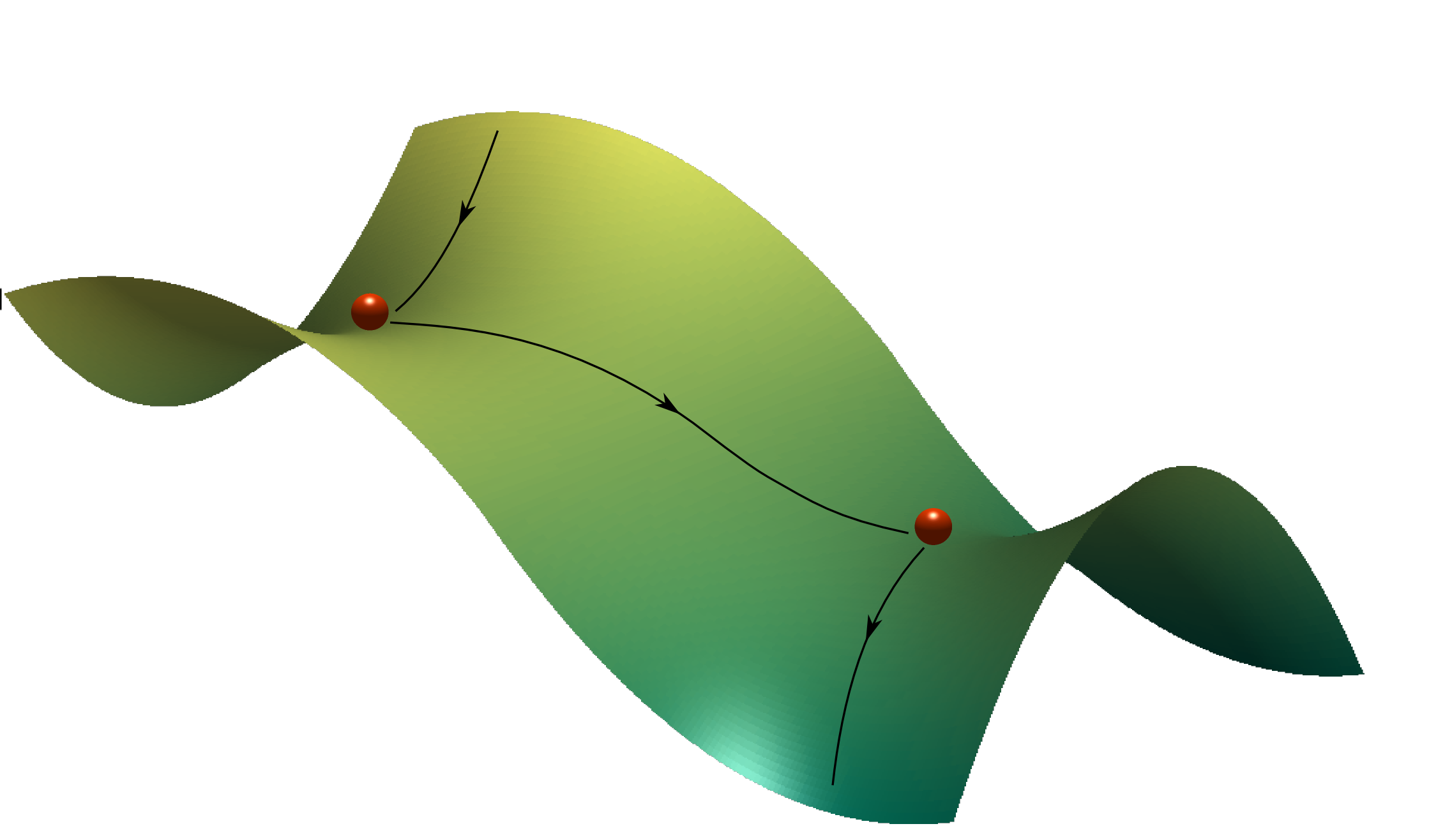}\hspace*{0.2cm}
\includegraphics[height=1.0in]{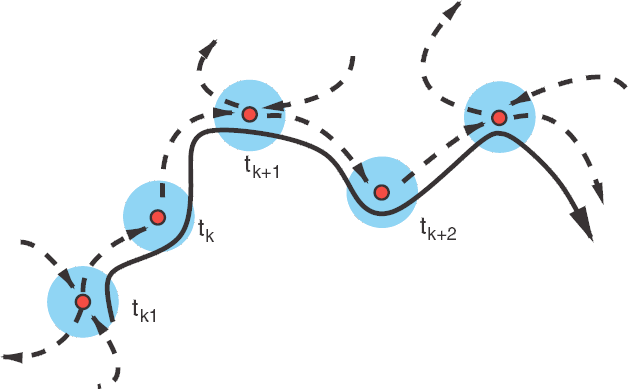}}
\caption{A heteroclinic channel is a sequence of metastable states (i.e., saddles) organized in such a way that the unstable manifold of one saddle is owned as the stable manifold of the following saddle (left panel). The channel can be (structurally) stable (i.e., attractive for any point in its vicinity); this is a common situation arising in a number of normal forms for a broad range of parameters.}\label{fig:shc}
\end{figure}

Since the time spent within a saddle vicinity is inversely proportional with the (logarithm of the) noise level (variance) \cite{Kif:81,StH:90}, the characteristic time of such a transient varies in a wide range.
In a stable heteroclinic sequence, the order of temporal winners is fixed and the noise is able to accelerate the process.
{\it Thus, the noise must be large enough to maintain the switching behavior at the desired rate (on average) and small enough to keep the heteroclinic nature of system on track, i.e. to maintain stability.} 

Embedding a structurally stable heteroclinic skeleton in the phase space (see Fig.~\ref{fig:shc}) results in a channel which routes the volume around it along the imposed sequence. Within this volume, the system behavior is reproducible with finite accuracy. Since the location 
of the saddles conveys input-specific information, which is activated the corresponding metastable states and their sequential order by the strength and the topology of the connections, then {\it the system becomes both noise-rejecting and input-sensitive (due to following the stimulus-specific channel) simultaneously.}
The key mechanism underlying the winnerless competition in brain is inhibition, which is known to exist in neural systems at micro- and macroscopic levels \cite{Aro:07,KUB:08,JLV:08,BKR:07}. 

There is substantial experimental support \cite{ABG:95,JFSM:07,RHL:08} that metastability and transient dynamics are the key phenomena underlying cortical processes and thus yield a better understanding of a dynamical brain.

\section{The Model}
\subsection{The Lotka-Volterra System}
As numerous results show, the Generalized Lotka-Volterra (GLV) model given by 
\begin{eqnarray}
\tau\frac{d}{dt}x_i(t) = x_i\left[\mu_i(E)-\sum_{i=1}^{n}\varphi_{ij}(E) x_i(t)\right] + x_i(t)\cdot\eta(t),\nonumber\\
i=1,\ldots,n.\label{eqn:GLV}
\end{eqnarray}
is playing the role of a normal form for the dynamical analysis of competition among $n$ agents (see, for example, \cite{ATHR:08}).
Here $x_i\geq 0$ is the $i$-th competing agent, $E$ is the input that captures all (known) external effects on the competition, $\tau$ is the time-constant, $\mu_i$'s are the increments that represent the resources available to the competitor $i$ to prosper, $\varphi_{ij}$ is the competition matrix, and $\eta(t)$ is a multiplicative noise perturbing the system. 

The basic GLV form, as introduced above, lacks a coupling term to account for the interaction between the two processes. A rigorous complement to GLV in our case requires more insight into how the individual dynamics of cognitive and emotional modes interact. Ecological models of competing modes are directly applicable in describing and prediction of temporal structure of emotional and cognitive functions, such as sequential learning, short-term memory, and decision making in a changing environment. 

The GLV equations can demonstrate a variety of dynamical behaviors. Among them we are particularly interested in the heteroclinic behavior, which arises when the parameters $\mu$ and $\varphi_{ij}$ are kept within certain ranges as formulated by in \cite{AZR:04}. 

\subsection{Ecological Model of Mental Dynamics}
We formulate the ecological model of the mental dynamics by three sets of variables. The first two are regarding the cognitive functions and the emotions, respectively:
\begin{eqnarray}
\tau_A\cdot\frac{d}{dt}A_i(t) & = & A_i(t)\left[\sigma_i(I,\mathbf{B})\cdot R_A -\right.\nonumber\\
& & \left. \sum_{j=1}^{N}\rho_{ij} A_j(t)\right] + A_i(t)\cdot\eta(t),\nonumber\\
&&\ i=1,\ldots,N, \label{eqn:cog}\\
\tau_B\cdot\frac{d}{dt}B_i(t) & = & B_i(t)\left[\zeta_i(S,\mathbf{A})\cdot R_B -\right.\nonumber\\
& & \left.\sum_{j=1}^{M}\xi_{ij} B_j(t)\right] + B_i(t)\cdot\eta(t),\nonumber \\
&&i=1,\ldots,M. \label{eqn:emot}
\end{eqnarray}
The nonnegative variables $A_i$ and $B_i$, as described above, correspond to the cognitive and emotional modes, the union of which are denoted by $\mathbf{A}$ and $\mathbf{B}$ respectively. The proposed model is merely a formulation of the competition within and among these two sets of modes. Both of these modes are open to the external world through the quantities $I$ and $S$, denoting the cognitive load and the stressor, respectively. The functions $\sigma_i(\cdot)$ and $\zeta_i(\cdot)$ are positive increments promoting the cognitive and the emotional processes, respectively. These terms' dependence on the cognitive and emotional modes enable the direct interaction between the two processes, which, depending on the exact form of these functions, can excite or inhibit each other.  These two coupled processes evolve on time scales determined by the parameters $\tau_A$ and $\tau_B$. Both processes are open to the brain noise, which appears as the multiplicative perturbation $\eta(t)$ in the equations.

The competition driving the mental processes in the joint emotion-cognition workspace is for the finite resources needed to carry out each process. Two basic types of resources are the energy (oxygen, glucose, etc.) and the information (attention and memory), both assumed to be supplied to the whole body of modes in constant amounts.

Working memory (WM) is a buffer for information over a short period of time. Experimental findings suggest that brain networks that are supported WM dynamics are strongly overlapping with networks that support attention dynamics. This overlap in function exists not only on an anatomical but also on a functional/dynamical level. Therefore we propose a model of attention-working memory informational recourses. Attention chooses from available information the items that are currently relevant to cognition including emotion, appraisal, and the coping strategy. Experiments suggest that any enhancing effect of attention can be explained by competitive interaction among attention modes that represent all of the considerable informational items.  

It is important to note that the latter resource is a product of the brain itself, and that, its availability may have different effects on each cognitive and emotional processes that evolve in parallel. The share that each process gets from the available resources is represented in the model by the two positive variables $R_A$ and $R_B$, which also compete among themselves separately, partitioning the unity:
\begin{eqnarray}
\frac{d}{dt}R_A(t) & = & R_A(t)\left[\sum_{i=1}^{N}A_i - \left(R_A(t)+\right.\right.\nonumber\\
&&\left.\left.\phi_A(I) R_B(t)\right)\right],\label{eqn:resrcA}\\
\frac{d}{dt}R_B(t) & = & R_B(t)\left[\sum_{i=1}^{M}B_i - \left(R_B(t)+\right.\right.\nonumber\\
&&\left.\left.\phi_B(S) R_A(t)\right)\right].\label{eqn:resrcB}
\end{eqnarray}
Two fixed parameters $\phi_A$ and $\phi_B$ determine this competition among two resources.

The model embodies competition in three different ways:
\begin{enumerate}
\item[(i)]{within the modes that make up the cognitive function or the emotion,}
\item[(ii)]{among the modes of the two processes,}
\item[(iii)]{via the resources available to each process.}
\end{enumerate}

\section{Simulations}
Here we give two examples of mental dynamics that illustrate the emergence of {\it negative} emotions and their interaction with cognitive processing. 

\begin{figure}
\centerline{\includegraphics[width=3in]{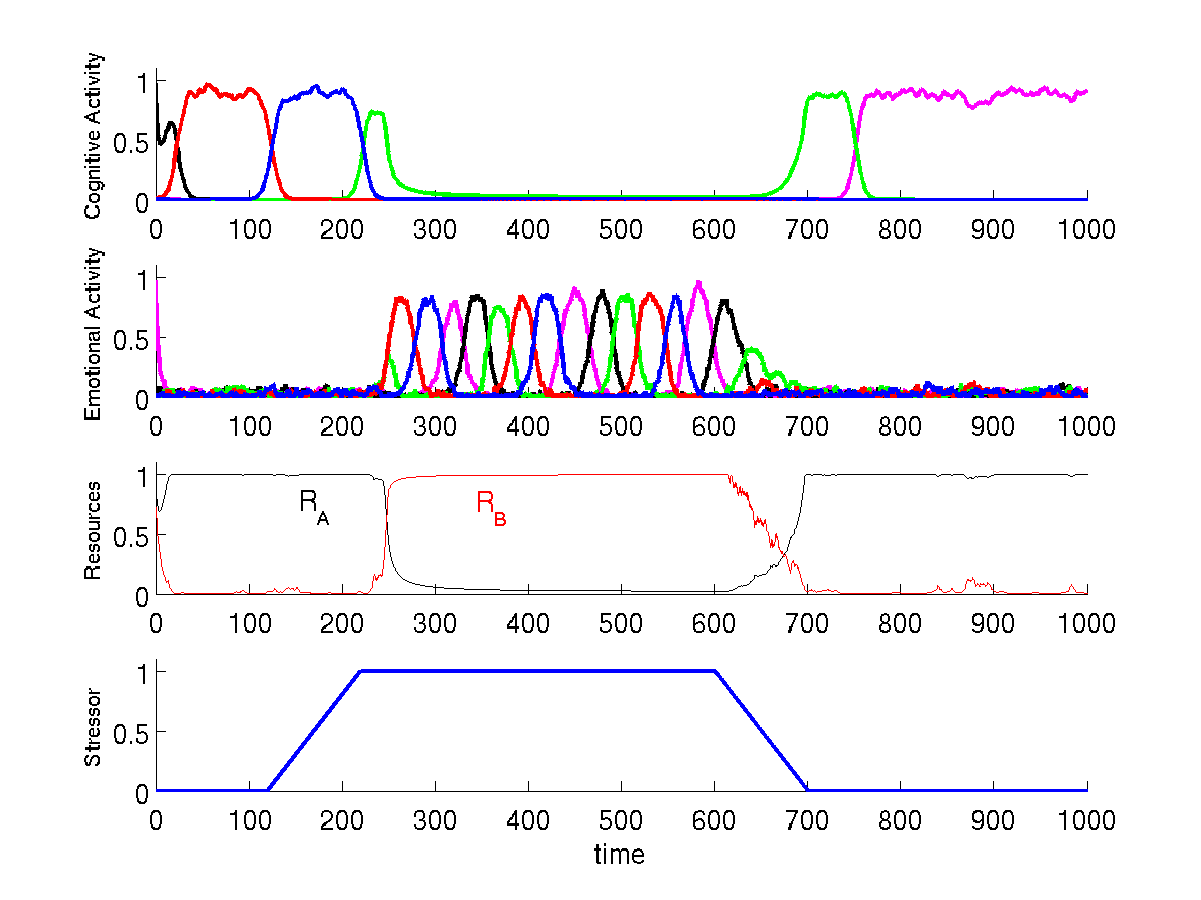}}
\caption{Simulation of the stressor-induced emotion-cognition interaction generated by the proposed ecological model. The bottom curve is the temporal profile of the stressor, which triggers the emotional activity depicted on the second row. Arousal of these emotional modes affect the ongoing cognitive activity negatively, as seen on the first row. This effect is due to two couplings between the cognitive and emotional processes: (i) the direct interaction encoded in the cognitive increments $\sigma$ (see text), and (ii) through the resource competition, whose trace is shown on the third row.}\label{fig:exp1}
\end{figure}

In the first example, an auxiliary stressor $S$ triggers these emotions, which in turn disrupts an ongoing cognitive sequence. Thus, the simulation demonstrates the feed-forward chain of events 
\[S\Rightarrow {\rm negative\ emotions}\Rightarrow {\rm cognitive\ disruption}\] 

Let us consider $N=5$ cognitive modes and $M=5$ emotional components. The multiplicative perturbation $\eta(t)$ is a white noise with variance $10^{-8}$ and $10^{-3}$ for the cognitive and the emotional dynamics, respectively, and the time constants are $\tau_A=\tau_B=20$.

Without loss of generality, we prescribed the finite heteroclinic sequence of saddles $e_1\rightarrow 2\rightarrow\cdots\rightarrow e_5$ for the emotional modes. The mode $e_5$ is a stable attractor (i.e., without any unstable manifold so that the system is confined to the vicinity of $e_5$ once it enters its domain of attraction). This state marks the terminal cognitive mode, such as the execution of a certain coping strategy, whereas the preceding modes denote the cognitive tasks that lead to this resulting activity. They could be named, for instance, as perception, appraisal, evaluation, and selection, in their order of appearance in the sequence.

The feasible values of $\rho_{ij}$ that can establish the desired heteroclinic skeleton in the $A$ network constitute a broad continuum in the parameter space. A set of sufficient conditions that determine a part of this region in the form of simple inequalities on $\sigma_i$ and $\rho_{ij}$ can be found in \cite{AZR:04}. Following these conditions, we set $\rho_{ii}=1.0$ for $i\in\{1,\ldots,5\}$, $\rho_{i-1,i}=1.5$ for $i\in\{2,\ldots,5\}$, $\rho_{i,i+1}=0.5$ for $i\in\{1,\ldots,4\}$, and $\rho_{ij}=\rho_{j-1,j}+2$ for $j\in\{2,3,4\}$ and $i\notin{j-1,j,j+1}$.

In this illustration, the five emotional modes were organized as a heteroclinic sequence, yet as a cyclic one by introducing $e_5\rightarrow e_1$ transition. We note that we do not necessarily name the emotional components individually, but interpret their mean activity as the degree of anxiety, a negative emotional state. In this respect, the precise dynamical quality of the emotional network is not of primary consideration in our design; for the sake of our illustrations, the emotional behavior could have been realized simply as a limit cycle, or as a strange attractor. The $\xi_{ij}$ was evaluated as done above for $\rho_{ij}$, yet taking into account the $e_5\rightarrow e_1$ transition, which results in $\xi_{5,5}=0.5$.

\begin{figure}
\centerline{\includegraphics[width=3in]{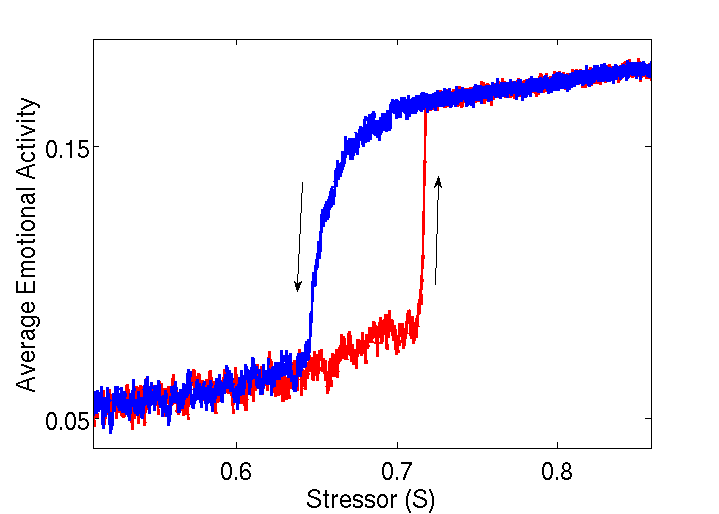}}
\caption{The emotional activity and the stressor level causing it have a hysteresis relation. The stressor magnitude that results in a given level of emotional activity is not unique and depends on whether the stressor is being modified in an increasing or decreasing direction. It can be shown that the mechanism responsible from this contrast is the indirect (resources) competition.}\label{fig:exp1_hyst}
\end{figure}

All five increments $\sigma_i$ in the cognitive process were modelled as $1-\sum_{i=1}B_i(t)$, i.e., inversely proportional to the total (negative) emotional activity. The increments $\zeta_i$ for the five emotional modes were considered as independent of the cognitive activity in this example; they were all equal to the externally applied stressor quantity $S$, which was assumed to be non-negative.

\begin{figure}
\centerline{\includegraphics[width=3in]{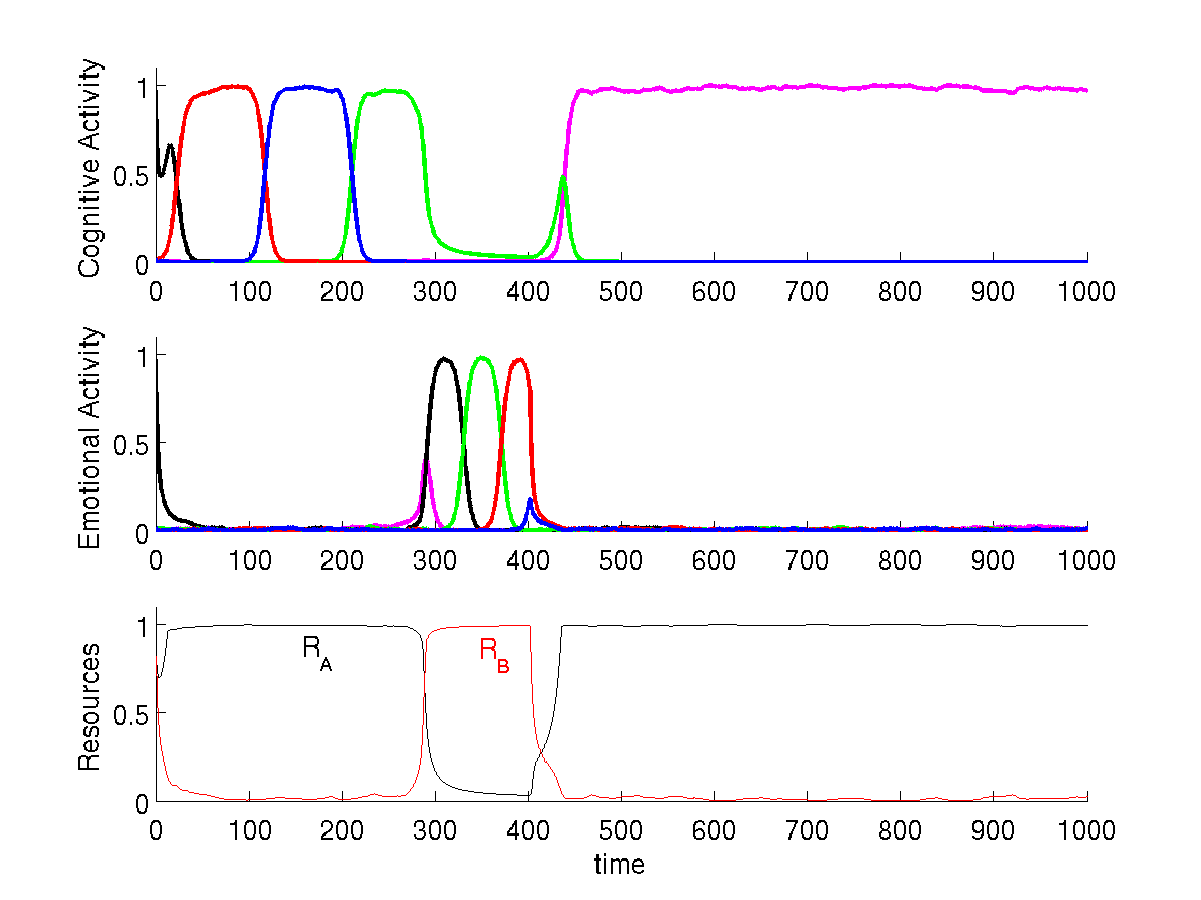}}
\caption{A self-induced emotion-cognition interaction as generated by the proposed ecological model. The interpretations are as in the previous figure. Here a certain cognitive mode, the $A_4$ denoted by the green curve, triggers the emotional activity, which suppresses the cognitive activity in return. The emotional activity is time-limited as encoded in $\xi$ (see text); the cognitive process returns to back its track after this period.}\label{fig:exp2}
\end{figure}

The resource competition $R_A$ vs. $R_B$ is regulated by the equations (\ref{eqn:resrcA}) and (\ref{eqn:resrcB}) with parameters $\phi_A=\phi_B=0.3$ and random initial conditions. 

With the selected parameters, the integration of the ordinary differential equations were performed by the Milstein approximation. The results shown in Fig. \ref{fig:exp1} were obtained. The figure illustrates the suppression of and delay in the cognition due to the emotional activity, which is induced by an external stressor. 

An interesting prediction that can be derived from the model is the contrast in the switching regimes of the total activity in the cognitive and the emotional network during the rising and decay periods of $S$. This can be better observed in Fig.~\ref{fig:exp1_hyst} which shown $\sum_{i=1}^5A_i$ versus $\sum_{i=1}^5{B_i}$ with the underlying $S$ color-coded.

The second example studies the self-induction of negative emotions by a certain cognitive mode, such as appraisal or self-evaluation.  In this case, we study the autonomous regime:
\[{\rm negative\ emotions}\rightleftharpoons{\rm cognitive\ disruption}.\]

\begin{figure}
\centerline{\includegraphics[width=3in]{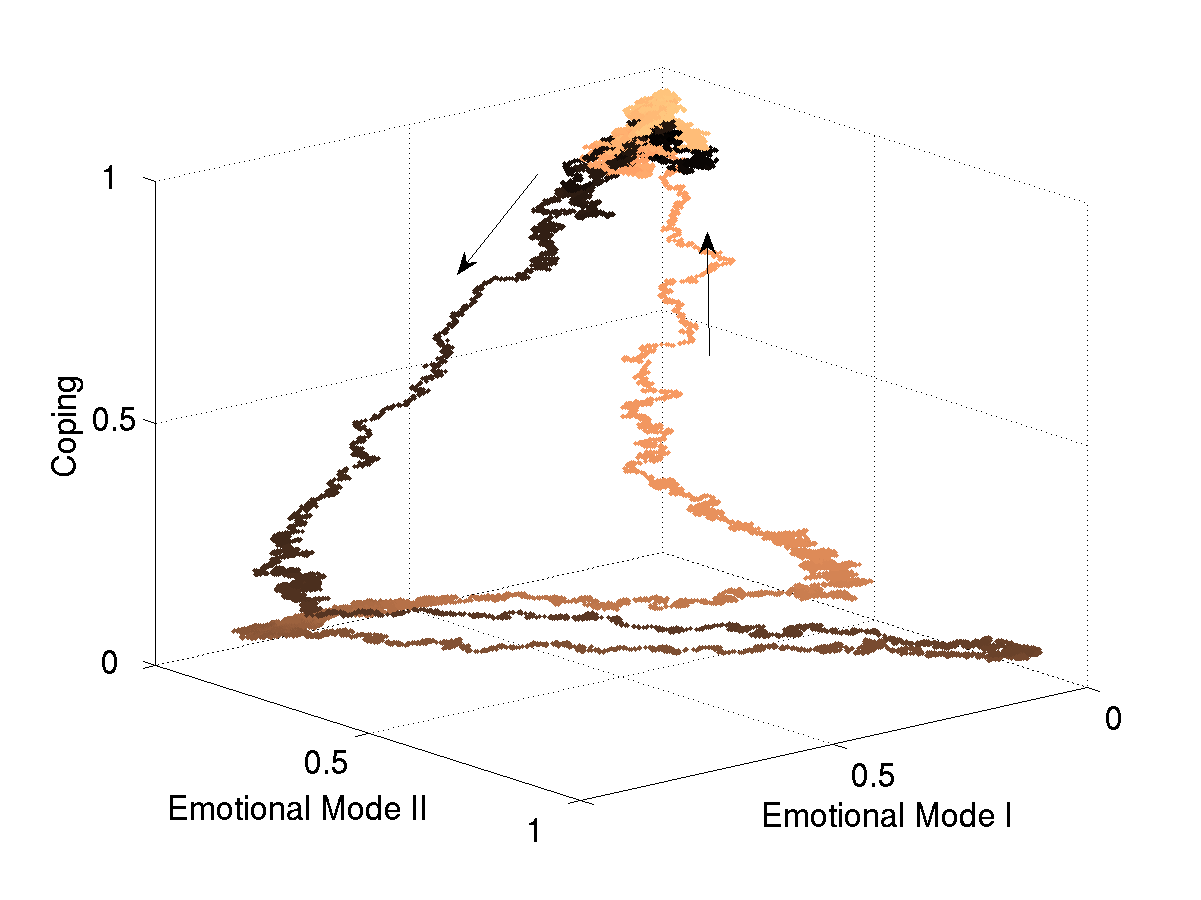}}
\caption{The change in the cognitive activity $\sum_{i=1}^5A_5$ with respect to the emotional activity which is separated into two axes. The color change from dark to light shows time arrow.}\label{fig:exp2_phase}
\end{figure}

The individual dynamical qualities (i.e., the mode orders and the heteroclinic skeletons) were the same as in the first example. All parameters and couplings were also retained, with the exception of the design of $\xi$. The increments for the emotional modes were assigned as $A_4$, the fourth cognitive mode and were all reset to zero after $200$ time units. Note that the triggering effect is {\it internal}, due to a specific cognitive mode. Figure~\ref{fig:exp2} shows a result of this simulation.  The phase portrait representing the dynamics of the total cognitive activity and the emotional modes is shown in Fig.~\ref{fig:exp2_phase}. 

\section{Conclusion}
We have presented here a unified theory of emotion and cognition that for the first time
considered their the dynamics of their mutual interactions in time. This theory accommodates
various components such as emotional-cognitive appraisal, generation of coping strategies, and
informational resources such as attention and working memory. The theory is rooted in general
dynamical principals: competition for brain resources including energetic and informational;
robustness, reproducibility and transitivity.

\section{Acknowledgments}
This work has been supported by U.S. Office of Naval Research through the grant ONR-N00014-07-1-0741.

\bibliography{refs}
\bibliographystyle{aaai}

\end{document}